\documentclass[%
 reprint,
superscriptaddress,
 amsmath,amssymb,
 aps,
]{revtex4-2}

\usepackage{graphicx}
\usepackage{dcolumn}
\usepackage{bm}
\usepackage{stmaryrd}
\usepackage{xcolor}

\usepackage[%
  colorlinks=true,
  urlcolor=blue,
  linkcolor=blue,
  citecolor=blue
]{hyperref}

\newcommand{\mbf}{\boldsymbol}
\newcommand{\me}{\hat{\boldsymbol{e}}}
\newcommand{\dd}{\mathrm{d}}

\newcommand{\pdif}[2]{\dfrac{\partial #1}{\partial #2}}

\begin{document}

\preprint{APS/123-QED}

\title{Geodesic Trapping and Escape of Active Particles on Curved Surfaces}

\author{Yuzhu Chen}
 \affiliation{Department of Mechanical and Aerospace Engineering, University of California San Diego, 9500 Gilman Drive, La Jolla, CA 92093, USA}
 
\author{Vishal P. Patil}%
 \affiliation{Department of Mathematics, University of California San Diego, 9500 Gilman Drive, La Jolla, CA 92093, USA}

\author{David Saintillan}
 \affiliation{Department of Mechanical and Aerospace Engineering, University of California San Diego, 9500 Gilman Drive, La Jolla, CA 92093, USA}


\begin{abstract}
Active particles on curved surfaces can become trapped along closed geodesics even without physical barriers. We show that escape from these geometric traps exposes a fundamental distinction between continuous and discrete reorientation. At high P\'eclet numbers, active Brownian particles escape efficiently through rotational diffusion, whereas run-and-tumble particles remain trapped much longer; at low P\'eclet numbers, both reduce to passive diffusion. Gaussian curvature controls escape by focusing or defocusing neighboring geodesics, producing distinct asymptotic scalings of the mean exit time.

\end{abstract}

\maketitle

Geometry plays a fundamental role in controlling and organizing the dynamics of active matter.
The interplay between geometry and self-propulsion in active matter systems can lead to rich phenomena \cite{bechinger2016active}, including boundary accumulation \cite{elgeti2015run}, trapping \cite{chepizhko2013diffusion}, and sorting \cite{volpe2011microswimmers,bruss2017curvature,iyer2023dynamics,schonhofer2022curvature,webb2026geometric}. 
Microscopically, these active matter systems typically consist of self-propelled particles, often classified by their orientation dynamics. Active Brownian particles (ABPs) can reorient gradually through rotational diffusion, whereas run-and-tumble particles (RTPs), such as bacteria, move in straight lines and reorient abruptly through tumbling events. Although strong confinement on the scale of the persistence length can distinguish ABPs from RTPs~\cite{solon2015active, khatami2016active}, these classes of self-propelled particles appear equivalent at large length scales in flat space~\cite{cates2013active}. However, curved surfaces are the natural setting for a broad range of biological active particle dynamics, from motor-driven cytoskeleton filaments on droplet interfaces or membrane vesicles \cite{giomi2014defect,ellis2018curvature,hsu2022activity} to migrating cells on curved substrates \cite{brandstatter2023curvature,happel2024coordinated} and morphogenesis \cite{maroudas2021topological,vafa2022active}. How curvature impacts this ABP--RTP equivalence remains poorly understood.

\begin{figure*}[!htbp]
\includegraphics[width=\textwidth]{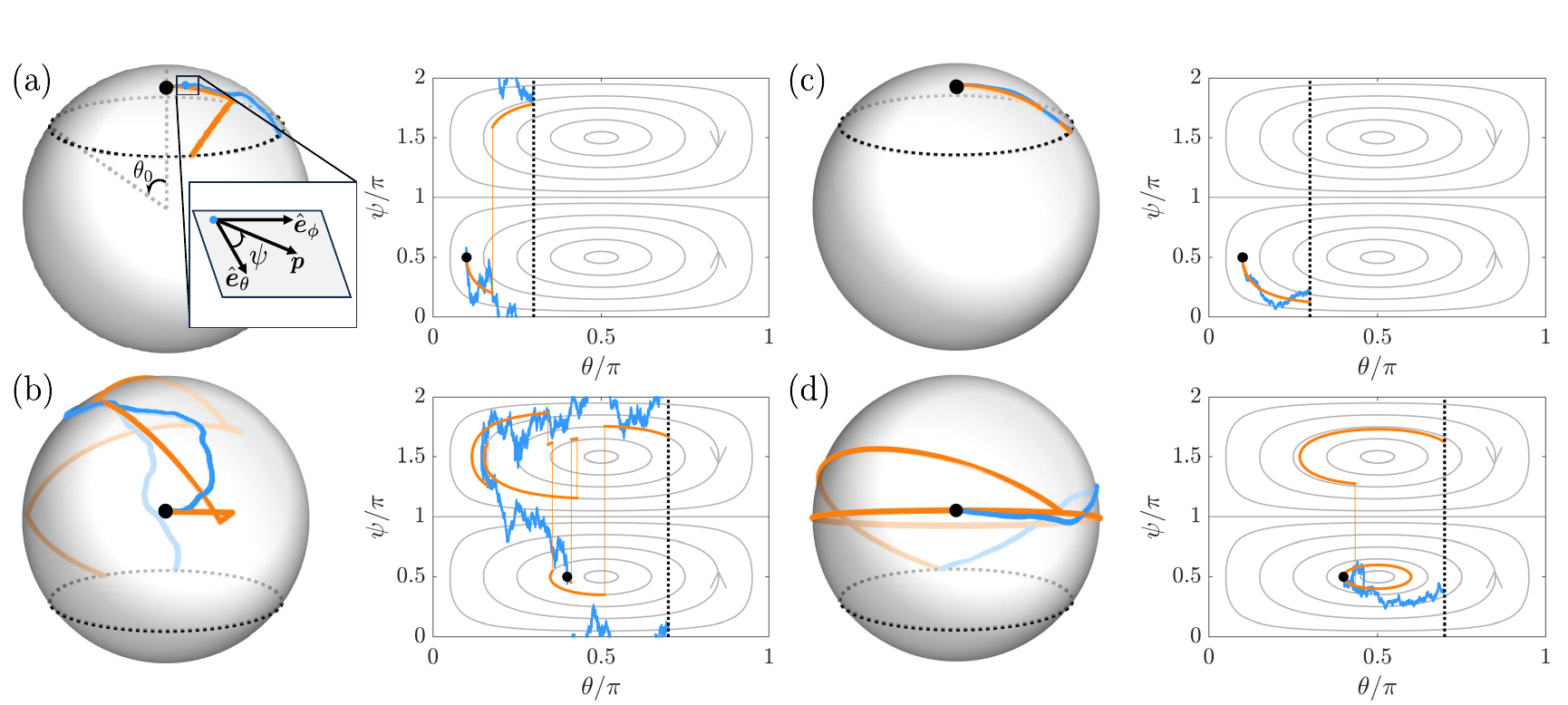}
\caption{\label{fig:fig1} Typical trajectories of ABPs (blue) and RTPs (orange) on a sphere and in $(\theta,\psi)$ phase space for $\mathrm{Pe}=1$ (a,b) and $\mathrm{Pe}=10$ (c,d). Black dots denote initial positions, dashed curves the exit boundaries, and gray curves the geodesic flow in $(\theta, \psi)$ space. In the top row (a,c), all geodesics intersect the exit boundary ($\theta_0<\pi/2)$, precluding deterministic trapping. In the bottom row (b,d), the domain contains closed geodesics, such as the equator ($\theta = \pi/2, \;\psi = \pi/2, 3\pi/2$), that never reach the exit boundary ($\theta_0>\pi/2$), giving rise to geodesic trapping of persistent particles. The inset in (a) illustrates the parametrization of the particle orientation in the local tangent plane.}
\end{figure*}

Transport along curved surfaces is mediated by Gaussian curvature.
In the passive case, curvature controls transport by modifying particle diffusivity \cite{faraudo2002diffusion,yoshigaki2007theoretically,castro2010brownian} and creating geometric diffusion barriers~\cite{molina2021diffusion}. In active systems, curvature couples to particle motion by modifying the orientations of the self-propelling particles. In particular, Gaussian curvature governs the convergence and divergence of neighboring geodesics, leading to effects analogous to gravitational lensing and tidal forces \cite{mackay2026emergent}. It can also disrupt flocking by changing the separation and relative orientation of neighboring particles, resulting in dispersion in flocking even with strong alignments \cite{mackay2026emergent,shibata2026geometric}.

In this Letter, we demonstrate that curvature can break ABP--RTP equivalence, thus producing statistically different active matter systems that explore space in distinct ways. This discrepancy emerges from the first passage time statistics for active particles.
Focusing on ABPs and RTPs, we combine Monte Carlo simulations, the numerical solutions of the mean exit time equations, and asymptotic analysis to study their exit dynamics on various curved geometries. 
On curved surfaces, geodesics can form closed orbits that trap deterministic particles even in the absence of physical barriers, which is referred to as geodesic trapping in the following text. 
Stochastic reorientation enables particles to escape geodesic traps, but highly depends on the reorientation mechanism when the persistence length is large. 
At large P\'eclet numbers, the continuous rotational diffusion of ABPs allows them to escape more efficiently, while RTPs remain trapped for longer times due to discrete tumbling events. 
We further show that, in the large $\text{Pe}$ limit, positive and negative Gaussian curvature-induced geodesic focusing and scattering can lead to distinct scaling laws for the maximum mean exit time of ABPs.
For RTPs, by contrast, curvature only affects the prefactor while leaving the scaling exponent unchanged.

We consider a self-propelled particle constrained to a fixed curved surface $\mathcal{M}$, with position $\mbf{x}\in\mathcal{M}$ and orientation $\mbf{p}=\cos\psi\,\me_1+\sin\psi\,\me_2\in T_{\mbf{x}}\mathcal{M}$, where $\me_{1,2}$ form a local orthonormal tangent frame and $\psi\in[0,2\pi)$. In the absence of stochastic reorientation, the particle follows a geodesic according to
\begin{align}
\dot{\mbf{x}} &= v_0\mbf{p}, \label{eq:dxdt}\\
\dot{\psi} &= -v_0p^\alpha\Omega_\alpha, \label{eq:dpsidt}
\end{align}
where $v_0$ is the constant self-propulsion speed and $\Omega_\alpha=\me_2\cdot\nabla_\alpha\me_1$ are the components of the spin connection \cite{kamien2002geometry,do2012differential,nakahara2018geometry}. The second equation compensates for rotation of the local frame, ensuring parallel transport of $\mbf{p}$ on the unit tangent bundle of $\mathcal{M}$ \cite{do1992riemannian}.

For an ABP, the orientation additionally undergoes rotational diffusion in the local tangent plane \cite{fily2016active,apaza2018active,castro2018active},
\begin{equation}
\dot{\psi}=-v_0p^\alpha\Omega_\alpha+\sqrt{2D_r}\eta(t),
\end{equation}
where $\langle\eta(t)\rangle=0$ and $\langle\eta(t)\eta(t')\rangle=\delta(t-t')$. By contrast, an RTP follows geodesics deterministically between instantaneous tumbles, where the run times $t_r$ are exponentially distributed with probability density function $p(t_r)=\lambda e^{-\lambda t_r}$, and each tumble uniformly randomizes $\psi$. We choose the tumbling rate $\lambda=D_r$, so that ABPs and RTPs have identical orientational correlation times and persistence lengths $\ell_p=v_0/D_r$. In both models, the P\'eclet number $\mathrm{Pe}=\ell_p/L=v_0/(D_rL)$ measures the persistence length relative to a characteristic geometric length scale $L$.

\begin{figure*}[!htbp]
\includegraphics[width=\textwidth]{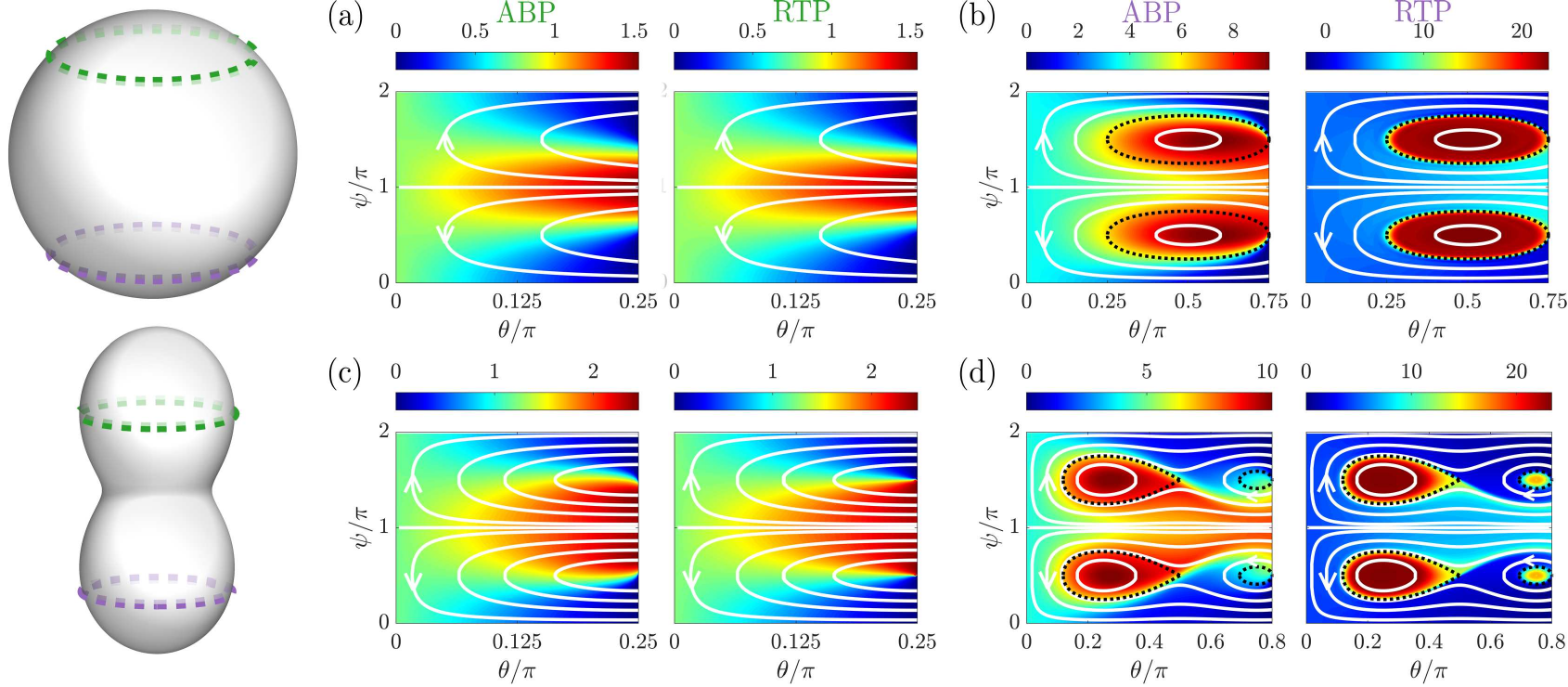}
\caption{\label{fig:fig2} Mean exit time $\tau$ in $(\theta,\psi)$ space for ABPs and RTPs at $\mathrm{Pe}=10$, obtained from the numerical solutions of Eqs. \eqref{MET:ABP} and \eqref{MET:RTP}. Each row compares ABPs (left) and RTPs (right). White solid curves show the geodesic flow, and black dashed curves delimit regions of geodesic trapping. Results are shown for a sphere with (a) $\theta_0=\pi/4$ and (b) $\theta_0=3\pi/4$, and for an hourglass shape of diameter $3L$ (see SM for details \cite{Supplement}) with (c) $\theta_0=\pi/4$ and (d) $\theta_0=0.8\pi$. The surface geometry and exit boundaries for each case are depicted in the left panel of each row. See SI \cite{Supplement} for additional results at other values of $\mathrm{Pe}$ and $\theta_0$. }
\end{figure*}

We first consider a sphere of radius $L$ parametrized by the standard spherical coordinates $(\theta,\phi)$, with an absorbing boundary along the latitudinal circle at $\theta = \theta_0$. Typical ABP and RTP trajectories are shown on the sphere and in $(\theta,\psi)$ phase space in Fig.~\ref{fig:fig1}. 
The nature of escape and corresponding exit times depend crucially on the location of the boundary. 
For $\theta_0<\pi/2$, every geodesic intersects the exit boundary, precluding deterministic trapping [Figs.~\ref{fig:fig1}(a,c)]. 
For $\theta_0>\pi/2$, by contrast, the domain contains closed geodesics that never reach the boundary, as seen from the gray streamlines in phase space [Figs.~\ref{fig:fig1}(b,d)]. 
Deterministic particles initialized on these geodesics therefore remain trapped indefinitely.

Stochastic reorientation provides a  mechanism for escape from geodesic trapping. 
At low $\text{Pe}$ [Figs.~\ref{fig:fig1}(a,b)], rapid reorientation allows both ABPs and RTPs to move readily between geodesics, leading to rapid escape on comparable time scales. At high $\mathrm{Pe}$, however, geodesic trapping exposes a fundamental distinction between continuous and discrete reorientation. Without trapping, when the exit boundary is at $\theta_0 < \pi/2$ [Fig.~\ref{fig:fig1}(c)], both ABPs and RTPs escape nearly ballistically. 
However, for $\theta_0>\pi/2$ [Fig.~\ref{fig:fig1}(d)], rotational diffusion allows ABPs to drift continuously across geodesics and out of the trapped region, whereas RTPs remain confined to a closed geodesic until a tumble places them on an escaping trajectory, resulting in much longer exit times. 

We quantify these differences through the mean exit time $\tau(\mbf{x},\psi)$, defined as the expected time to first reach the absorbing boundary starting from position $\mbf{x}\in \mathcal{M}$ and orientation $\psi$. We obtain $\tau$ by numerically solving the corresponding Kolmogorov backward equations \cite{thiffeault2018exit} on the curved surface. In dimensionless form, the ABP mean exit time satisfies
\begin{align}
   p^\alpha \partial_{\alpha} \tau - p^\alpha\Omega_\alpha \pdif{\tau}{\psi} + \frac{1}{\text{Pe}} \pdif{^2\tau}{\psi^2} = -1, \label{MET:ABP}
\end{align}
whereas for RTPs, 
\begin{align}
    & p^\alpha \partial_{\alpha} \tau - p^\alpha\Omega_\alpha \pdif{\tau}{\psi} - \frac{\tau}{\text{Pe}} + \frac{1}{2\pi \text{Pe}}\int_0^{2\pi}\!\!\! \tau(x^\alpha, \psi')\dd \psi' = -1, \label{MET:RTP}
\end{align}
where the boundary conditions are taken as $\tau = 0$ along the outflow boundaries. 
A derivation of these equations and boundary conditions, as well as a validation of their numerical solution against Monte Carlo simulations are presented in the Supplementary Material \cite{Supplement}.

The mean exit time function captures how curvature-mediated reorientation dynamics leads to statistically distinct classes of active particles. This distinction depends on the geodesic structure of the manifold. When the exit boundary is located in the upper hemisphere [Fig. \ref{fig:fig2}(a)], the mean exit time distributions of ABPs and RTPs are qualitatively similar. 
Since every geodesic intersects the exit boundary, there is no trapped region and the maximum mean exit time for both ABPs and RTPs corresponds to trajectories starting at the boundary, $\theta(0) = \theta_0$, and pointing into the domain, $\psi(0) = \pi$ [Fig. \ref{fig:fig2}(a)]. 
By contrast, when the exit boundary is in the lower hemisphere [Fig. \ref{fig:fig2}(b)], there exists a trapped region of closed geodesics that do not intersect the exit boundary. In this regime, the continuous reorientation of ABPs allows them to escape the trapped region more quickly than RTPs, which exhibit substantially larger mean exit times. In the $(\theta, \psi)$ plane, the boundary of the trapped region can be predicted exactly from conservation of the quantity $p_\phi = \sin\theta \sin\psi$ along geodesics, which can be interpreted as angular momentum. Geodesics satisfying $|p_\phi|>|\sin\theta_0|$ cannot reach the exit boundary, yielding the separatrix 
$|\sin\theta_0| = |\sin\theta \sin\psi|$ 
shown by the black dashed curves in Fig.~\ref{fig:fig2}(b). Points in this region therefore satisfy $\pi - \theta_0 \leq \theta, \psi \leq \theta_0$.
The enclosed deterministic trapped regions indeed coincide with regions of large $\tau$ for both ABPs and RTPs (see movie S1 of the SI \cite{Supplement} for a video showing particle simulations).

The geodesic structure also controls escape on more complex surfaces, where the same trapping mechanism can lead to breakdown in ABP--RTP equivalence. For the hourglass-shaped surface considered in the lower row of Fig. \ref{fig:fig2}, the region of negative Gaussian curvature around $\theta = \pi/2$ is associated with saddle structures in the geodesic flow and corresponding separatrices in phase space. 
For $\theta_0=\pi/4$ [Fig. \ref{fig:fig2}(c)], the domain contains no trapped regions, and the mean exit-time fields resemble those of the spherical cap, with no significant difference between ABPs and RTPs. 
When both lobes of the hourglass are contained within the domain, as for $\theta_0=0.8\pi$ [Fig. \ref{fig:fig2}(d)], multiple trapped regions emerge, corresponding to closed geodesics confined to each lobe. As on the sphere, RTPs exhibit much longer exit times than ABPs within these trapped regions (see movie S2 \cite{Supplement}).

\begin{figure}[t]
\includegraphics[width=\columnwidth]{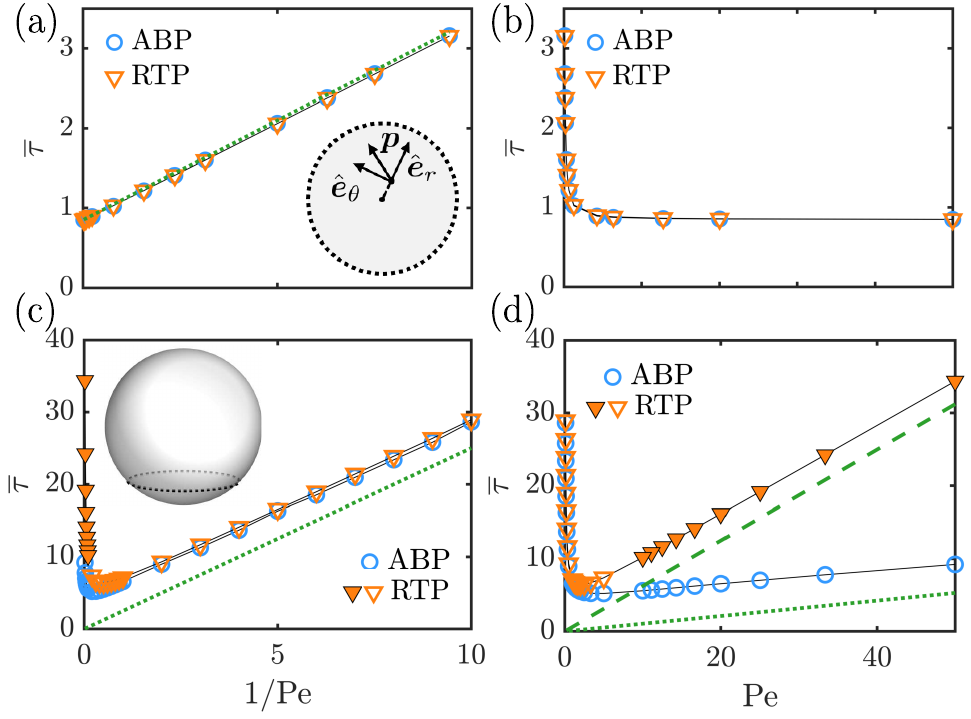}
\caption{\label{fig:fig3} Configuration-averaged mean exit time $\bar{\tau}$ for ABPs and RTPs on (a,b) a flat disk and (c,d) a spherical cap with $\theta_0=0.75\pi$. Results are plotted versus $1/\mathrm{Pe}$ in (a,c) and $\mathrm{Pe}$ in (b,d). Open symbols denote numerical solutions of the mean exit-time equations, and filled symbols denote particle-based simulations. Green dashed lines show asymptotic predictions derived in the SI \cite{Supplement}. }
\end{figure}

We next examine how geodesic trapping modifies the dependence of the mean exit time on particle dynamics. To this end, we consider the configuration-averaged mean exit time $\bar{\tau}$, obtained by averaging $\tau(\mbf{x},\psi)$ uniformly over initial positions and orientations in the domain. 
Figure~\ref{fig:fig3} compares $\bar{\tau}$ for ABPs and RTPs on a flat disk of radius $L$, which contains no trapped geodesics, and on a spherical domain containing closed geodesics, both plotted as functions $\mathrm{Pe}$, with $1/\mathrm{Pe}$ used to highlight the low-P\'eclet asymptotics. At low $\mathrm{Pe}$, ABPs and RTPs behave diffusively and thus become indistinguishable on both geometries, with $\bar{\tau}\sim \mathrm{Pe}^{-1}$. Indeed, in that limit, an asymptotic expansion of the mean exit time of the form $\tau = \tau_0/\text{Pe} + \tau_1 + \text{Pe}\,\tau_2 + \cdots $ yields 
\begin{align}
    \frac{1}{2} \Delta_{LB} \tau_0 = -1  \label{eq:tau0eq}
\end{align}
for the leading-order term in both models, corresponding to passive diffusion with effective diffusivity $\mathrm{Pe}/2$, consistent with classical results \cite{bechinger2016active}. At low $\mathrm{Pe}$, geometry thus enters only through the Laplace-Beltrami operator $\Delta_{LB}$ in Eq.~(\ref{eq:tau0eq}), which depends on the metric and determines the prefactor of the common $\mathrm{Pe}^{-1}$ scaling \cite{faraudo2002diffusion,yoshigaki2007theoretically,castro2010brownian}. 

At large $\mathrm{Pe}$, however, the presence of closed geodesics fundamentally alters the escape dynamics. On the flat disk, every geodesic intersects the boundary, and therefore $\bar{\tau}$ for both ABPs and RTPs approaches the same finite ballistic limit of $8/3\pi$ in Fig.~\ref{fig:fig3}(b). On the spherical domain, by contrast, $\bar{\tau}$ is nonmonotonic and diverges as $\mathrm{Pe}\to \infty$ [Fig.~\ref{fig:fig3}(d)]. 
The nonmonotonicity reflects a crossover between two escape mechanisms: increasing persistence initially accelerates diffusive exploration of the domain, but at large $\mathrm{Pe}$ suppresses the reorientation required to escape from the trapped regions. The divergence is much stronger for RTPs: a trapped RTP must wait an $O(\mathrm{Pe})$ time for a tumble to redirect it onto an escaping geodesic, yielding $\bar{\tau}_{\mathrm{RTP}}\sim \mathrm{Pe}$. The prefactor, corresponding to the slope of the thick green dashed curve in Fig.~\ref{fig:fig3}(d), can be obtained analytically from the Poisson statistics of tumbling events as detailed in the Supplementary Material \cite{Supplement}. ABPs, on the other hand, escape through continuous rotational diffusion along the orientational coordinate $\psi$ and exhibit a much weaker divergence with $\mathrm{Pe}$, reflecting a fundamentally different mechanism for escape from geodesic trapping. We estimate the maximum escape time as $\max(\tau)\sim l^2 \text{Pe}/2$, where $l = \theta_0 - \pi/2$ is the maximum half-width of the trapped region along $\psi$. The prefactor in Fig.~\ref{fig:fig3}(d) is averaged by assuming the mean exit time to be constant within the trapped region and zero outside.

\begin{figure*}[t]
\includegraphics[width=\textwidth]{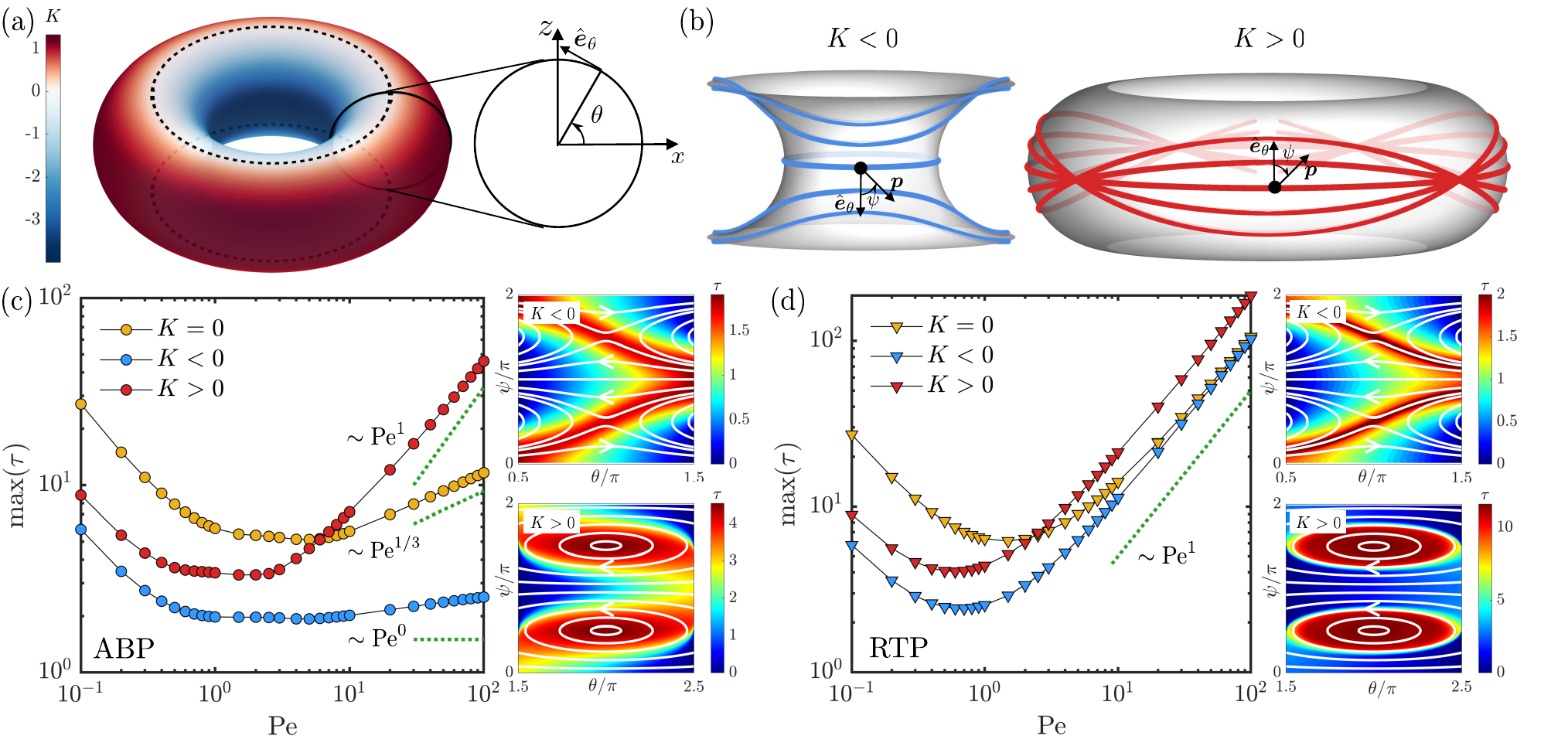}
\caption{Mean exit time on toroidal patches with different Gaussian curvatures, obtained from the numerical solutions of Eqs. \eqref{MET:ABP} and \eqref{MET:RTP}. (a) Schematic of a torus colored by Gaussian curvature $K$, with blue and red patches corresponding to regions of negative (inner) and positive (outer) curvature, respectively. The black dotted curve shows the exit boundary for each patch. (b) Geodesic defocusing for $K<0$ and focusing for $K>0$, illustrated by initially parallel geodesics. (c,d) Maximum mean exit time $\max (\tau)$ versus  $\text{Pe}$ for (c) ABPs and (d) RTPs on patches with $K=0$, $K<0$ and $K>0$. Green dotted lines indicate large-$\mathrm{Pe}$ scalings. Contour plots show the mean exit time fields in  $(\theta,\psi)$ space for $K<0$ and $K>0$ at $\mathrm{Pe}=5$, where white curves are streamlines of the geodesic flow. \label{fig:fig4} }
\end{figure*}

In contrast to positive Gaussian curvature, which focuses geodesics, regions of negative and zero Gaussian curvature inhibit and reverse this effect. Together, these effects can be used to control and sort ABPs and RTPs in the large P\'eclet limit. Geodesic focusing is governed by the Jacobi equation~\cite{do2016differential}, which relates the Gauss curvature $K$ to the distance $J(s)$ between neighboring, arc-length-parametrized geodesics, through the ODE $J'' = -KJ$. Accordingly, spheres, which have constant positive Gaussian curvature, lead to trapping of ABPs and RTPs as demonstrated above. On the other hand, generic surfaces exhibit mixed curvatures [Fig. \ref{fig:fig4}(a)], where the diverging geodesics of negatively curved regions can enhance escape.

We demonstrate signed-curvature control of ABP and RTP dynamics by examining their behavior on negatively and positively curved torus strips with major radius $L$ and minor radius $rL$
[Fig. \ref{fig:fig4}(b)], as well as a flat strip with $K=0$. In the case of  ABPs in Fig. \ref{fig:fig4}(c), modifying curvature leads to distinct scaling regimes for the exit time $\tau$. When $K>0$, converging geodesics lead to a larger scaling exponent, with $\max(\tau) \sim\text{Pe}^1$ approximately, analogous to the spherical case. In the $K<0$ strip, continuously reorienting ABPs can rapidly move to diverging geodesics, leading to a nearly constant $\max(\tau)$ as $\text{Pe}$ increases.
In the flat strip case ($K=0$), we identify an intermediate scaling regime. The mean exit time equation on the flat strip reads
\begin{align}
    \cos\psi \pdif{\tau}{x} + \frac{1}{\text{Pe}}\pdif{^2\tau}{\psi^2} = -1. 
\end{align}
In the limit of large $\text{Pe}$, the scaling of $\tau$ can be obtained by introducing the stretched variables $\delta \psi = \psi - \pi/2 = \text{Pe}^{-\alpha} \Psi$, since there exists a boundary layer along $\psi = \pi/2$. 
Balancing the terms for advection and rotational diffusion gives $\alpha = 1/3$, while balancing the resulting leading-order term with the constant $-1$ yields $\max(\tau) \sim \smash{\text{Pe}^{1/3}}$, in good agreement with numerical simulations [Fig. \ref{fig:fig4}(c)]. In contrast, this curvature-dependent scaling is not observed in RTPs [Fig. \ref{fig:fig4}(d)], since the escape of RTPs is primarily governed by the run time between tumbling events, which scales as $\text{Pe}^1$. 

Gaussian curvature determines the structure of the geodesic flow, and hence controls both the area of the trapped regions in phase space and the prefactor of the exit-time scaling for RTPs. On the negatively-curved torus strip with dimensionless small radius $r$, the separatrices satisfying $|(1 + r\cos\theta)\sin\psi| = 1-r$ that approach the saddle points correspond to ridges with large mean exit time. In the positive curvature case, the closed orbits satisfying $|(1 + r\cos\theta)\sin\psi| = 1$ are tangent to the exit boundary and enclose finite trapped regions in phase space where the mean exit time is large. 
These phase-space structures suggest that the geodesic flow can organize active transport on curved surfaces through fixed points, separatrices, and closed orbits, analogous to the Hamiltonian phase-space structure that separates bounded and unbounded orbits of microswimmers in vortical flows \cite{tanasijevic2022microswimmers}. 

In summary, we have shown that curved surfaces can fundamentally alter active particle transport through geodesic trapping and escape. Closed geodesics provide purely geometric traps, without physical barriers, whose escape dynamics expose a fundamental distinction between continuous and discrete reorientation. While ABPs and RTPs become equivalent in the diffusive or flat limits, their exit times differ markedly at high persistence in curved domains: rotational diffusion allows ABPs to drift across trapped geodesics, whereas RTPs remain confined until a tumble redirects them onto an escaping trajectory. Gaussian curvature further controls escape through the focusing and defocusing of neighboring geodesics, producing distinct asymptotic scalings of the mean exit time. Geodesic structure thus determines the global landscapes of trapped and escaping trajectories, and provides a mechanism by which intrinsic geometry can distinguish active dynamics that are equivalent in flat space.
This mechanism could be relevant to target-searching of self-propelled particles \cite{wang2016target,rupprecht2016optimal} and has potential application to the sorting of active particles with different reorientational dynamics by designing surface geometries without external chemical signals or fluid flows.

\nocite{*}

\bibliography{apssamp}

\end{document}


\maketitle

\vspace{-1.3cm}
\begin{center}
\footnotesize $^1$ \textit{Department of Mechanical and Aerospace Engineering, \\
University of California San Diego, 9500 Gilman Drive, La Jolla, CA 92093, USA}\\
$^2$ \textit{Department of Mathematics, 
University of California San Diego, 9500 Gilman Drive, La Jolla, CA 92093, USA}
\end{center}

\setcounter{figure}{0}
    \renewcommand{\thefigure}{S\arabic{figure}}

\section{Derivation of the SDEs of the ABP}
Following \cite{apaza2018active,castro2018active}, we reformulate the dynamics of an ABP using Cartan's moving frame method. 
For simplicity, we neglect translational diffusion, and the particle moves at a constant self-propulsion speed $v_0$ along its instantaneous direction $\bp$ following
\begin{align}
\dot{\mx} = v_0 \bp. 
\end{align}
The orientational dynamics is governed by
\begin{align}
    \nb_{\dot{\mx}}\, \bp = \sqrt{2D_r}(\mbf{I}-\mbf{NN})\cdot[\bp\times(\mbf{\eta}\times \bp)],
\end{align}
where $\nb_{\dot{\mx}}\bp$ denotes the covariant derivative of $\bp$ with respect to $\dot{\mx}$, $\mbf{I}$ is the identity matrix in $\mathbb{R}^3$, $\mbf{N}$ is the surface normal, and $\mbf{\eta}$ is the Gaussian white noise in $\mathbb{R}^3$. 
Consider a set of orthonormal local frame $\{\me_1,\me_2\}\in T_{\mx} \mathcal{M}$, the orientation can be expressed using the local frame and an angle $\psi\in [0,2\pi)$. 
Therefore, the LHS can be written as
\begin{align}
     \nb_{\dot{\mx}}\, \bp & = \nb_{\dot{\mx}} (\cos\psi \, \me_1 + \sin\psi \, \me_2) \\
     & = -\dot \psi \sin\psi \,\me_1 + 
     \cos\psi\,\nb_{\dot{\mx}}\, \me_1 + 
     \dot \psi \cos\psi \,\me_2 + 
     \sin\psi\,\nb_{\dot{\mx}}\,\me_2. 
\end{align}
Introducing the spin-connection $\Omega_\alpha=\me_2\cdot\nabla_\alpha\me_1$ \cite{kamien2002geometry,do2012differential,nakahara2018geometry}, we have
\begin{align}
    & \nb_{\dot{\mx}}\,\me_1 = v_0 p^\alpha \nb_{\alpha} \me_1 = v_0 p^\alpha \Omega_{\alpha} \me_2, \quad \nb_{\dot{\mx}}\,\me_2 = v_0 p^\alpha \nb_{\alpha} \me_2 = - v_0 p^\alpha \Omega_{\alpha} \me_1.
\end{align}
Therefore, after rearranging the RHS we obtain
\begin{align}
    (\dot{\psi} + v_0 p^\alpha \Omega_\alpha)\bp^\perp = \sqrt{2D_r}(\bp^\perp \bp^\perp)\cdot \mbf{\eta}, 
\end{align}
where $\bp^\perp = -\sin\psi \,\me_1 + \cos\psi\,\me_2$ is the 90 degree counterclockwise rotation of $\bp$. 
Taking the inner product with $\bp^\perp$ on both sides gives
\begin{align}
    \dot{\psi} = -v_0 p^\alpha \Omega_\alpha + \sqrt{2D_r} \bp^\perp \cdot \mbf{\eta}. 
\end{align}

The noise term $\bp^\perp\cdot \mbf{\eta}$ can be shown to be equivalent to a one-dimensional white noise, so the rotational dynamics is given by
\begin{align}
    \dot{\psi} = -v_0 p^\alpha \Omega_\alpha + \sqrt{2D_r} \eta(t).
\end{align}

\section{Derivation of the mean exit time equation}
From the stochastic equation of the ABP, the Fokker-Planck equation can be written as
\begin{align}
    \pdif{\Psi}{t} = - \nb_\alpha (v_0 p^\alpha \Psi) + \pl_\psi (v_0 p^\alpha \Omega_\alpha \Psi) + D_r \pl_\psi^2 \Psi, 
\end{align}
where $\Psi(\mx,\bp,t)$ is the probability density function. 

We introduce the mean exit time, that is, the average time for particles to exit the boundary of the spatial domain $\mathcal{M}$ starting from a position initially at $(\mx_0,\psi_0,t_0)$ with the initial condition
\begin{align}
    \Psi(\mx,\psi,t_0|\mx_0,\psi_0,t_0) = \delta(\mx,\mx_0)\delta(\psi,\psi_0),
\end{align}
where $\Psi(\mx, \psi,t)$ is the probability density distribution function on the configuration space $\Omega = \mathcal{M}\times S^1$. 
Here we assume that the entire spatial boundary of the domain is an exit region $\partial_e\Omega$, where the particles are allowed to leave the domain. 
In the absence of translational diffusion, the Fokker-Planck equation is purely advective in space, and the exit regions can be divided into two parts according to the direction of the flows. 
The particles can only leave the domain if they hit the boundary regions with orientation $\bp$ satisfying $\bp \cdot \mbf{n} > 0$, denoted as $\partial \Omega_e^-$, called the outflow region, where $\mbf{n}$ is the outward normal on the exit boundary. 
For the regions where $\bp\cdot \mbf{n} <0$ (denoted by $\partial \Omega_e^+$, the inflow region), the inflows bring particles into the domain. 
We assume that there is no particle appearing at the boundary due to inflow, so the boundary condition of the Fokker-Planck equation is required to be
\begin{align}
\Psi = 0 \text{ at } \partial \Omega_e^+,
\end{align}
and is only applied in the inflow region. 

To derive the mean exit time equation, we use the standard adjoint method \cite{thiffeault2018exit,tanasijevic2022microswimmers}, summarized as follows and with extension to mean exit time problem on curved surfaces for self-propelled particles. 
We start our discussion with general expressions for ABPs, where the forward Fokker-Planck equation defined on $\Omega$ can be written as
\begin{align}
    \partial_t \Psi + \mathcal{L}\Psi = 0,
\end{align}
with the operator 
\begin{align}
    \mathcal{L}\Psi \coloneqq \nb \cdot (\bu\Psi - \mbf{D} \cdot \nb \Psi),
\end{align}
where $\nb$ is an operator defined in $\Omega$ by
\begin{align}
    \nb \coloneqq (\nbp ,\; \pdif{}{\psi}),
\end{align}
and 
\begin{align}
\bu = (v_0\bp,\; -v_0p^\alpha \Omega_\alpha), \quad 
    \mbf{D} = \begin{pmatrix} 0 & 0 & 0 \\
    0 & 0 & 0\\
    0 & 0 & D_r
    \end{pmatrix}.
\end{align}
We can further define the adjoint of $\mathcal{L}$ with respect to the inner product
\begin{align}
    \langle f,g\rangle = \int_{\Omega} f g \,\dd V = \int_{\mathcal{M}}\int_{S^1} f(\mx,\psi)g(\mx,\psi) \,\dd \mx \,\dd \psi
\end{align}
by
\begin{align}
     \langle f,\mathcal{L}g\rangle = \langle \mathcal{L}^\ast f,g\rangle. 
\end{align}
By the definition of $\mathcal{L}g$, after performing integration by parts, the right-hand side of the above equation can be written as
\begin{align}
    \langle f, \mathcal{L} g\rangle = &\int_{\partial \Omega} [f\mn\cdot(\bu g - \mbf{D}\nb g) + \mn\cdot(g \mbf{D}\cdot \nb f)]\,\dd S \notag \\
    &- \int_{\Omega} g[\bu\cdot \nb f + \nb\cdot (\mbf{D}\cdot \nb f)]\,\dd V, 
\end{align}
so the adjoint operator can be defined as
\begin{align}
\mathcal{L}^\ast f = - \mbf{u}\cdot \nb f - \nb \cdot (\mbf{D}\cdot \nb f).
\end{align}
For the boundary terms to vanish, from the boundary conditions for the forward problem, the adjoint boundary conditions should be
\begin{align}
    & f = 0 \text{ at } \partial \Omega_e^- .
\end{align}
Applying the Chapman-Kolmogorov equation and the adjoint operator \cite{thiffeault2018exit}, we can obtain the Kolmogorov backward equation with respect to $(\mx_0,\psi_0,t_0)$:
\begin{align}
    - \partial_{t_0} \Psi + \mathcal{L}^\ast_{\mx_0,\psi_0,t_0}\Psi = 0 
\end{align}
with the adjoint boundary conditions above. 

Following the derivations in \cite{thiffeault2018exit},  the mean exit time equation can be written as
\begin{align}
    & \mathcal{L}^\ast_0 \tau = 1, \label{eq:MET}
\end{align}
with boundary condition
\begin{align}
    & \tau = 0 \text{ at } \partial \Omega_e^-. 
\end{align}

For ABPs, the mean exit time equation can be written as
\begin{align}
    v_0 p^\alpha \partial_\alpha \tau - v_0 p^\alpha \Omega_\alpha \pdif{\tau}{\psi} + D_r \pdif{^2\tau}{\psi^2} = -1,
\end{align}
and for RTPs, the mean exit time equation reads
\begin{align}
    &  v_0 p^\alpha \cdot \partial_\alpha \tau  -v_0 p^\alpha \Omega_\alpha \pdif{\tau}{\psi} - \lambda \tau + \frac{\lambda}{2\pi}\int_0^{2\pi} \tau(x^\alpha, \psi')\dd \psi' = -1.
\end{align}

\section{Governing equations in coordinates}
\paragraph{Sphere}
We use the standard spherical coordinates to parametrize the surface, where the position vector is
\begin{align}
    \mbf{x}(\theta,\phi) = \begin{pmatrix} \sin\theta \cos\phi \\ \sin\theta \sin\phi \\ \cos\theta \end{pmatrix},
\end{align}
the covariant basis is 
\begin{align}
    \mg_\theta = \begin{pmatrix} \cos\theta \cos\phi \\ \cos\theta \sin\phi \\ -\sin\theta \end{pmatrix} = \hat{\mbf{e}}_\theta,\quad \mg_\phi = \begin{pmatrix} -\sin\theta\sin\phi \\\sin\theta\cos\phi \\ 0 \end{pmatrix} = \sin\theta \hat{\mbf{e}}_\phi.
\end{align}
We take $\hat{\me}_1 = \etheta$ and $\hat{\me}_2 = \ephi$. 
The mean exit time equations are
\begin{align}
   & \cos\psi \pdif{\tau}{r} - \frac{\sin\psi}{r}\pdif{\tau}{\psi} + \frac{1}{\text{Pe}} \pdif{^2\tau}{\psi^2} = -1, \\
   & \cos\psi \pdif{\tau}{r} - \frac{\sin\psi}{r}\pdif{\tau}{\psi} - \frac{1}{\text{Pe}} \tau + \frac{1}{2\pi\text{Pe}}\int_0^{2\pi}  \tau(x^\alpha, \psi')\dd \psi' = -1
\end{align}
for ABPs and RTPs, respectively. 

\paragraph{Hourglass}
For the hourglass with radius $R(\theta)$, the position vector reads
\begin{align}
    \mbf{x}(\theta,\phi) = R(\theta)\begin{pmatrix}
    \sin\theta \cos\phi \\
    \sin\theta \sin\phi \\
    \cos\theta
    \end{pmatrix},
\end{align}
and the covariant basis is
\begin{align}
    \mg_\theta = R_{\theta} \hat{\me}_r + R \hat{\me}_{\theta}, \quad 
    \mg_\phi = R\sin\theta \hat{\me}_{\phi},
\end{align}
and the set of orthonormal frames can be constructed as
\begin{align}
& \hat{\me}_1 = \frac{1}{\sqrt{R^2 + R_\theta^2}}\mg_\theta, \quad \hat{\me}_2 = \frac{1}{R\sin\theta}\mg_\phi. 
\end{align}

The mean exit time equations are
\begin{align}
    & \frac{\cos\psi}{\sqrt{R^2 + R_\theta^2}}\pdif{\tau}{\theta} -  \frac{\sin\psi}{\sqrt{R_\theta^2 + R^2}}(\frac{R_\theta}{R} + \cot\theta) \pdif{\tau}{\psi} + \frac{1}{\text{Pe}} \pdif{^2 \tau}{\psi^2} = -1,\\
    & \frac{\cos\psi}{\sqrt{R^2 + R_\theta^2}}\pdif{\tau}{\theta} -  \frac{\sin\psi}{\sqrt{R_\theta^2 + R^2}}(\frac{R_\theta}{R} + \cot\theta) \pdif{\tau}{\psi} - \frac{1}{\text{Pe}} \tau + \frac{1}{2}\pi\text{Pe}\int_0^{2\pi}  \tau(x^\alpha, \psi')\dd \psi'  = -1 ,
\end{align}
for ABPs and RTPs, respectively. 

In the main text, the dimensionless radius is taken as $R(\theta) = 1 + 0.5\cos(2\theta)$, which has been scaled by the reference radius as a characteristic length. 

\paragraph{Torus}
The length scale of the torus is given by its major radius $R$. 
With the dimensionless major radius $1$ and minor radius $r$, it can be parametrized by 
\begin{align}
    \mbf{x}(\theta,\phi) = \begin{pmatrix}
    (1 + r\cos\theta)\cos\phi \\
    (1 + r\cos\theta) \sin\phi \\
    r \sin\theta
    \end{pmatrix},
\end{align}
The covariant basis is
\begin{align}
    \mg_{\theta} = \begin{pmatrix}
    -r\sin\theta\cos\phi \\
    -r\sin\theta\sin\phi \\
    r \cos\theta
    \end{pmatrix},
    \quad
    \mg_{\phi} = \begin{pmatrix}
    -(1 + r\cos\theta)\sin\phi \\
    (1 + r\cos\theta) \cos\phi \\
    0
    \end{pmatrix},
\end{align}
and the set of orthonormal frames can be constructed as
\begin{align}
& \hat{\me}_1 = \frac{1}{r}\mg_\theta, \quad \hat{\me}_2 = \frac{1}{1 + r\cos\theta}\mg_\phi. 
\end{align}

The mean exit time equations are 
\begin{align}
    & \frac{\cos\psi}{r} \pdif{\tau}{\theta} + \frac{\sin\theta\sin\psi}{1 + r \cos\theta}\pdif{\tau}{\psi} + \frac{1}{\text{Pe}}\pdif{^2 \tau}{\psi^2} = -1,\\
    & \frac{\cos\psi}{r} \pdif{\tau}{\theta} + \frac{\sin\theta\sin\psi}{1 + r \cos\theta}\pdif{\tau}{\psi} - \frac{1}{\text{Pe}} \tau + \frac{1}{2\pi\text{Pe}}\int_0^{2\pi}  \tau(x^\alpha, \psi')\dd \psi'  = -1 .
\end{align}
for ABPs and RTPs, respectively. 

In the main text, the minor radius is taken as $r=0.5$. 

\section{Comparison between particle simulations and numerical solutions of MET equations}

\begin{figure}[h]
\centering
\includegraphics[width=0.6\textwidth]{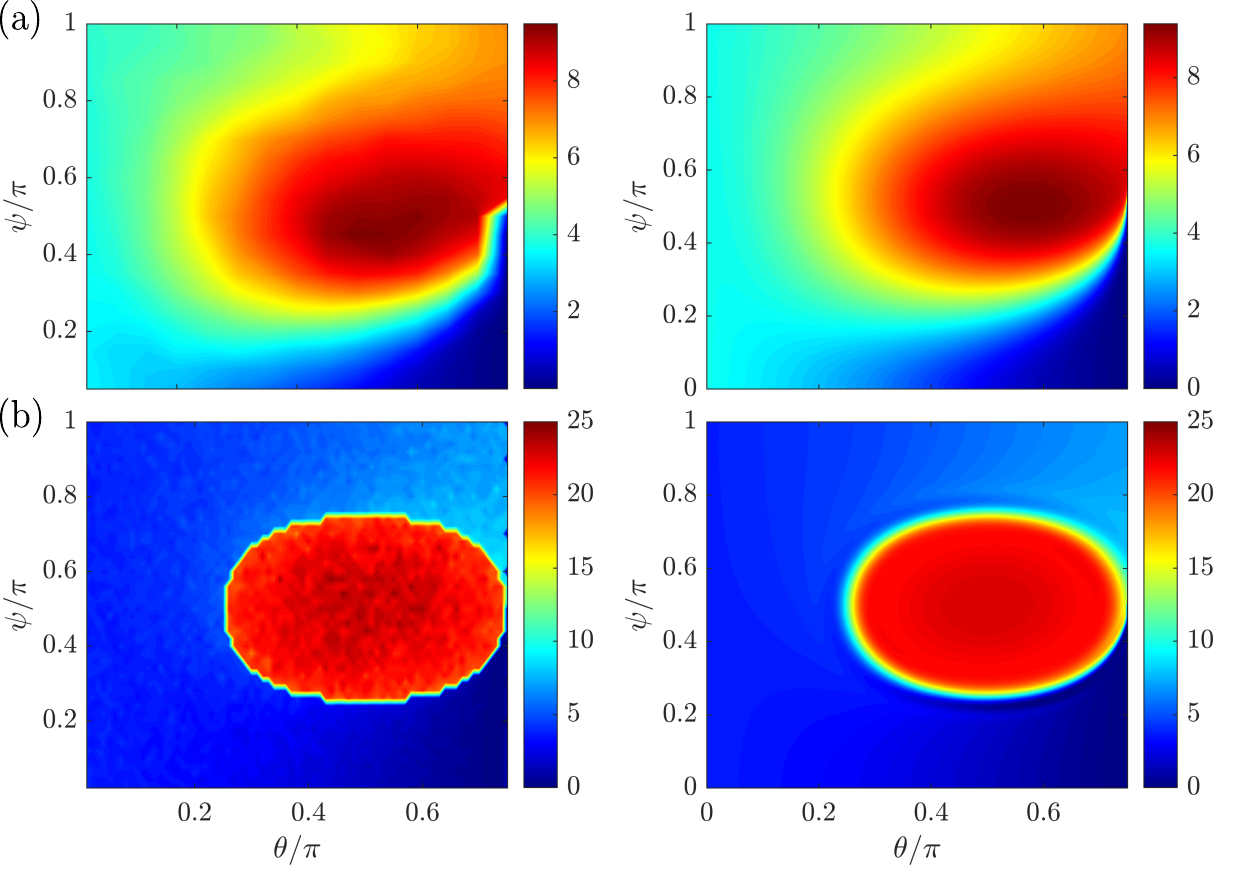}
\caption{\label{fig:sphere-compare} Comparison between particle simulations and numerical solutions of the MET equations of (a) ABPs and (b) RTPs on a spherical cap with $\theta_0 = 0.75\pi$ and $\text{Pe} = 10$. Left: particle simulations. Right: numerical solutions of the MET equations. Particle simulations are averaged over 1000 particles. \vspace{0.2cm}}
\end{figure}

\begin{figure}[h!]
\centering
\includegraphics[width=0.8\textwidth]{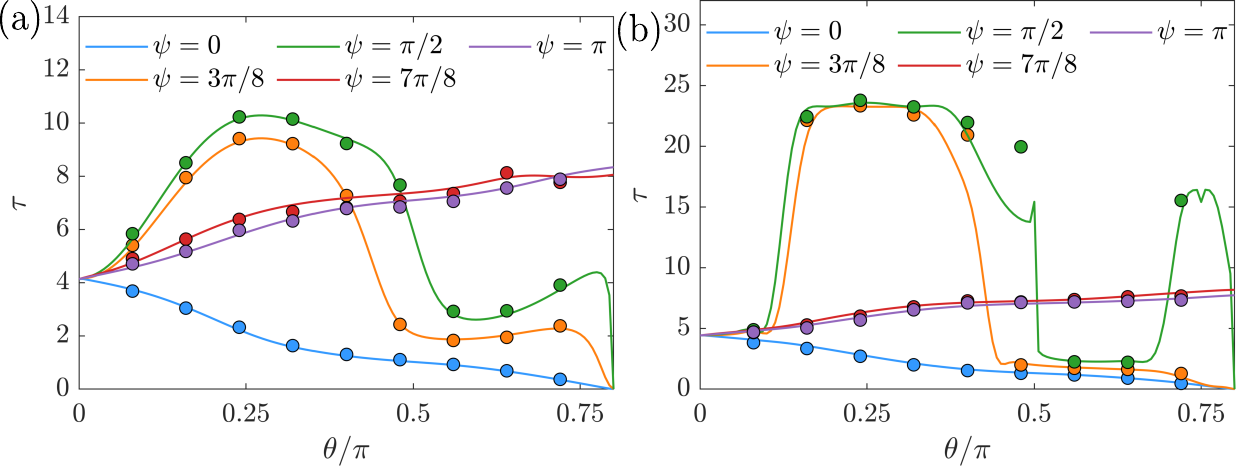}
\caption{\label{fig:hourglass-compare} Comparison between mean exit times of (a) ABPs and (b) RTPs on a hourglass cap with $\theta_0 = 0.8\pi$ and $\text{Pe} = 10$. Dots: particle simulations. Lines: numerical solutions of the MET equations. Particle simulations are averaged over 1000 particles. }
\end{figure}

A comparison of the mean exit time fields from Monte Carlo simulations and numerical solutions of the continuous mean exit time equations in the case of a sphere with $\theta_0=0.75\pi$ and $\mathrm{Pe}=10$ is shown in Fig.~\ref{fig:sphere-compare} for both ABPs and RTPs. In particle simulations, the mean exit time was obtained by averaging the exit time over an ensemble of 1000 particles initialized in each $(\theta,\psi)$ combination with fixed $\phi$.  A similar comparison on an hourglass shape with is shown in Fig.~\ref{fig:hourglass-compare}, where the mean exit time is plotted as a function of $\theta$ for different $\psi$'s. Excellent quantitative agreement is found in both cases, with the largest discrepancies occurring in the sharp gradients that are present near the boundary of the trapped regions.

\section{Asymptotic analysis in the small-$\text{Pe}$ limit}
\paragraph{ABP} In the small $\text{Pe}$ limit, we introduce the asymptotic expansion of the mean exit time as
\begin{align}
    \tau = \frac{1}{\text{Pe}}\tau_0 + \tau_1 + \text{Pe}\, \tau_2 + \cdots. 
\end{align}
Inserting the expansion into the mean exit time equation for ABPs, at $O(\text{Pe}^{-2})$
\begin{align}
    \pdif{^2\tau_0}{\psi^2} = 0,
\end{align}
which gives
\begin{align}
    \tau_0 = \tau_0(\mx)
\end{align}
owing to the periodicity in $\psi$. 
For the next order $O(\text{Pe}^{-1})$, we have the PDE
\begin{align}
    p^\alpha \nabla_\alpha \tau_0 + \pdif{^2\tau_1}{\psi^2} = 0.
\end{align}
With the periodic boundary conditions in $\psi$, the solution has the form 
\begin{align}
    \tau_1(\mx,\psi) =  p^\alpha \partial_\alpha \tau_0 + f(\mx). 
\end{align}
The next order equation at $O(1)$ reads
\begin{align}
    p^\alpha \partial_\alpha \tau_1 - p^\alpha \Omega_\alpha \pdif{\tau_1}{\psi} + \pdif{^2\tau_2}{\psi^2} = -1. 
\end{align}
Inserting the expression for $\tau_1$, we obtain
\begin{align}
    p^\alpha p^\beta \nb_\alpha \nb_\beta \tau_0 + p^\alpha \partial_\alpha f + \pdif{^2\tau_2}{\psi^2} = -1. 
\end{align}
Averaging the above equation over $\psi$ and using periodicity in $\psi$, we arrive at the leading order equation
\begin{align}
    \frac{1}{2}\Delta_{LB}\tau_0 = -1. 
\end{align}

\paragraph{RTP} Inserting the expansion into the mean exit time equation for RTPs, at $O(\text{Pe}^{-2})$
\begin{align}
    -\tau_0 + \frac{1}{2\pi}\int_0^{2\pi}\tau_0(\mbf{x},\psi) \dd \psi = 0,
\end{align}
which implies that
\begin{align}
    \tau_0 = \tau_0(\mx).
\end{align}
For the next order $O(\text{Pe}^{-1})$, we have
\begin{align}
    p^\alpha \pl_\alpha \tau_0 - \tau_1 + \frac{1}{2\pi}\int_0^{2\pi} \tau_1 \dd \psi = 0.
\end{align}
With the periodic boundary conditions in $\psi$, the solution has the form 
\begin{align}
    \tau_1(\mx,\psi) =  p^\alpha \partial_\alpha \tau_0 + \frac{1}{2\pi}\int_0^{2\pi}\tau_1 \dd \psi. 
\end{align}
The next order equation at $O(1)$ reads
\begin{align}
    p^\alpha \partial_\alpha \tau_1 - p^\alpha \Omega_\alpha \pdif{\tau_1}{\psi} - \tau_2 + \frac{1}{2\pi}\int_0^{2\pi} \tau_2 \dd \psi = -1. 
\end{align}
Inserting the expression for $\tau_1$, we obtain
\begin{align}
    p^\alpha p^\beta \nb_\alpha \nb_\beta \tau_0 + p^\alpha \partial_\alpha f - \tau_2 + \frac{1}{2\pi}\int_0^{2\pi} \tau_2 \dd \psi = -1. 
\end{align}
Averaging the above equation over $\psi$ and using  periodicity in $\psi$, we arrive at the same leading-order equation as for ABPs,
\begin{align}
    \frac{1}{2}\Delta_{LB}\tau_0 = -1. 
\end{align}

\section{Asymptotic analysis in the large-$\text{Pe}$ limit on the spherical cap}
\paragraph{ABP} We estimate the scaling of the mean exit time of ABPs on the spherical cap as follows. 
We first estimate $\tau_{\text{max}}$ as $l^2 \text{Pe}/2$, where $l = \theta_0 - \pi/2$ is the max half width of the trapped region along $\psi$. 
We then assume that within the trapped region the mean exit time is equal to $\tau_{\text{max}}$, and outside the trapped region is zero, which gives an area fraction factor of $2(1-\sin\theta_0)/(1-\cos\theta_0)$. 
Averaging over phase space the yields the slope of high-$\mathrm{Pe}$ scalings for ABPs shown in Fig. 3(d) of the main text.  

\paragraph{RTP}
We then estimate the scaling of the mean exit time of RTPs on a spherical cap in the limit of large $\text{Pe}$. 
We only consider the case where $\theta_0>\pi/2$, that is, there exist trapped regions.  
Given an exit boundary $\theta_0$, the geodesic phase space can be separated into two regions: the trapped region and the exit region. 
For particles uniformly distributed on the spherical cap, the probability density is 
\begin{align}
    p(\theta|\theta<\theta_0) = \frac{\frac{1}{2}\sin\theta}{\int_0^{\theta_0}\frac{1}{2}\sin\theta \dd \theta} = \frac{\sin\theta}{1-\cos\theta_0}. 
\end{align}
Given $\theta$, the probability that the particles are located within the trapped region is
\begin{align}
    P(\text{trapped}|\theta) = \begin{cases}
    1-\frac{2}{\pi}\arcsin\left(\frac{\sin\theta_0}{\sin\theta}\right),\quad \pi-\theta_0<\theta<\theta_0,\\
    0, \quad 0<\theta<\pi-\theta_0.
    \end{cases}
\end{align}
Hence, on the spherical cap, the probability for a particle to be initially trapped is
\begin{align}
    P_0(\text{trapped}|\theta<\theta_0) &= \int_{\pi-\theta_0}^{\theta_0} P(\text{trapped}|\theta)\; p(\theta|\theta<\theta_0) \dd \theta \notag \\
    &= \frac{2(1-\sin\theta_0)}{1-\cos\theta_0}. 
\end{align}
Particles initially located in the exit region will follow geodesic flow and exit, and therefore their contribution to the mean exit time is of order $O(1)$. 
For particles initially in the trapped region, since the tumbling time is exponentially distributed with mean $\text{Pe}$, before their first tumble, the contribution of the initially trapped particles to the mean exit time is
\begin{align}
    \tau_0 = \text{Pe} P_0. 
\end{align}

The first tumbling event can lead to four possible outcomes in the geodesic phase space: 

\noindent (i) A particle is initially trapped and tumbles into the trapped region. 

\noindent (ii) A particle is initially trapped and tumbles into the exit region. 

\noindent (iii) A particle is initially in the exit region and tumbles into the trapped region. 

\noindent (iv) A particle is initially in the exit region and tumbles into the exit region. 

The only case that can contribute to the mean exit time by an additional $O(\text{Pe})$ term is case (i). 
In case (ii), the particles can follow the geodesic flow and escape, so the contribution is O(1). 
In case (iii), since the probability for the particles to tumble before exit is $O(1/\text{Pe})$, with the subsequent trapping time $O(\text{Pe})$, the net contribution is $O(1/\text{Pe})\cdot O(\text{Pe}) = O(1)$. 
In case (iv), the probability for the particles to tumble before exit is $O(1/\text{Pe})$, and they can escape with $O(1)$, so the contribution is $O(1/\text{Pe})$. 

\begin{figure}[t]
    \centering
    \includegraphics[width=0.75\textwidth]{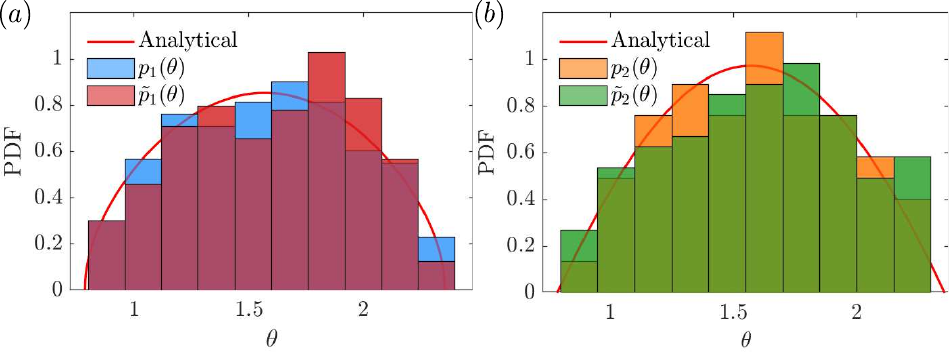}
    \caption{\label{fig:before after} Probability distribution of trapped RTPs before/after being advected by the geodesic flow. ($a$) Before the first tumble. Parameters: $\theta_0=0.75\pi$, $\text{Pe}=10$. ($b$) Before the second tumble.}  
\end{figure}

Therefore, we only focus on case (i), where the particle remains trapped after the first tumble. 
To estimate the probability density of the initially trapped particle after advection by the geodesic flow, we first estimate the probability density of the initially trapped particle immediately before its first tumble. 
A key assumption we make is that the flow preserves the $\theta$-marginal distribution of the initially trapped particle before the first tumble, from which 
\begin{align}
p_1(\theta| \text{initially trapped}) = \frac{\sin\theta\left(1-\frac{2}{\pi}\arcsin\left(\frac{\sin\theta_0}{\sin\theta}\right)\right)}{\int_{\pi-\theta_0}^{\theta_0} \sin\theta\left(1-\frac{2}{\pi}\arcsin\left(\frac{\sin\theta_0}{\sin\theta}\right)\right) \dd \theta}.
\end{align}
This assumption is validated numerically by particle simulations, as shown in Figure \ref{fig:before after}($a$), where $p_1(\theta)$ denotes the distribution before the flow and $\tilde{p}_1(\theta)$ denotes the distribution after being advected by the flow during the first run before the first tumble. 
After the first tumble, the conditional probability that the initially trapped particle remains trapped becomes
\begin{align}
     P_1(\text{trapped after first tumble}|\text{initially trapped}) 
    & = \int_{\pi-\theta_0}^{\theta_0} P(\text{trapped}|\theta) p_1(\theta| \text{initially trapped}) \dd \theta \notag \\
    &= \int_{\pi-\theta_0}^{\theta_0} \frac{\sin\theta}{2(1-\sin\theta_0)}\left(1-\frac{2}{\pi}\arcsin\left(\frac{\sin\theta_0}{\sin\theta}\right)\right)^2 \dd \theta. 
\end{align}
The contribution from waiting time between the first and second tumbling events is therefore
\begin{align}
    \tau_1 = \text{Pe} P_0P_1. 
\end{align}

For the second tumbling event, we can use the same argument as above, under the same assumption that the flow preserves the $\theta$-marginal distribution of the particles that are initially trapped and remain trapped after the first tumble, which is validated by particle simulations in Figure  \ref{fig:before after}($b$), where $p_2(\theta)$ denotes the distribution before the flow and $\tilde{p}_2(\theta)$ denotes the distribution after being advected by the flow during the second run before the second tumble. 
This distribution can be written as
\begin{align}
& p_2(\theta| \text{initially trapped and trapped after first tumble}) \notag \\
&\qquad= \frac{p_1(\theta)\left(1-\frac{2}{\pi}\arcsin\left(\frac{\sin\theta_0}{\sin\theta}\right)\right)}{\int_{\pi-\theta_0}^{\theta_0} p_1(\theta)\left(1-\frac{2}{\pi}\arcsin\left(\frac{\sin\theta_0}{\sin\theta}\right)\right) \dd \theta} = \frac{p_1(\theta)\left(1-\frac{2}{\pi}\arcsin\left(\frac{\sin\theta_0}{\sin\theta}\right)\right)}{P_1}.
\end{align}
Therefore, the probability for a particle that is initially trapped and remains trapped after the first tumble to remain trapped after the second tumble is
\begin{align}
    & P_2(\text{trapped after second tumble}|\text{initially trapped and trapped after first tumble}) \notag \\
    & \qquad= \int_{\pi-\theta_0}^{\theta_0} P(\text{trapped}|\theta) p_2(\theta| \text{initially trapped and trapped after first tumble}) \dd \theta \notag \\
    &\qquad = \int_{\pi-\theta_0}^{\theta_0} \frac{\sin\theta}{2(1-\sin\theta_0)P_1}\left(1-\frac{2}{\pi}\arcsin\left(\frac{\sin\theta_0}{\sin\theta}\right)\right)^3 \dd \theta,
\end{align}
and its contribution is
\begin{align}
    \tau_2 = \text{Pe} P_0P_1P_2. 
\end{align}

Following the same procedure above, we can obtain the recursive expression
\begin{align}
    P_n = \int_{\pi-\theta_0}^{\theta_0} \frac{\sin\theta}{2(1-\sin\theta_0)P_1P_2 \cdots P_{n-1}}\left(1-\frac{2}{\pi}\arcsin\left(\frac{\sin\theta_0}{\sin\theta}\right)\right)^{n+1} \dd \theta,
\end{align}
and the contribution between the $n$-th tumble and the $(n+1)$-th tumble is
\begin{align}
    \tau_n = \text{Pe} P_0 P_1 \cdots P_n. 
\end{align}
In summary, the asymptotic approximation in the limit of $\text{Pe}\to \infty$ for RTPs on a spherical cap is
\begin{align}
\tau \approx \tau_0 + \tau_1 + \tau_2 + \cdots.
\end{align}
In the main text we keep six terms for the approximated slope in Fig. 3(d) of the main text. The integrals are calculated numerically.

\section{Mean exit time fields with varying $\text{Pe}$}

\begin{figure}[htbp]
    \centering
    \includegraphics[width=0.8\textwidth]{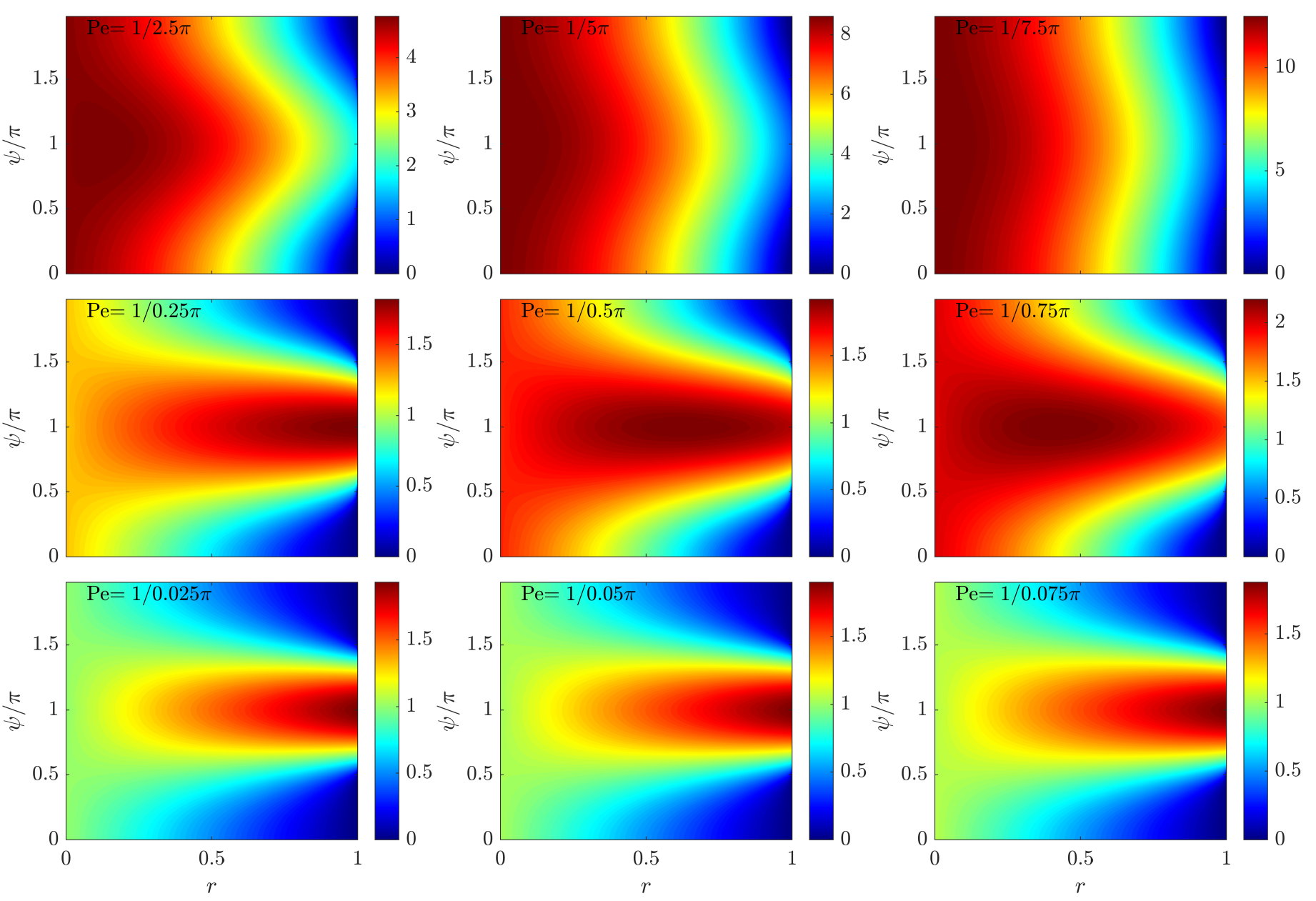}
    \caption{\label{fig:ABP disk MET contour} Contour plots of mean exit time of ABPs with different $\mathrm{Pe}$ in a flat disk. See the geometry in Fig. 3(a) of the main text.}  
\end{figure}

\begin{figure}[htbp]
    \centering
    \includegraphics[width=0.8\textwidth]{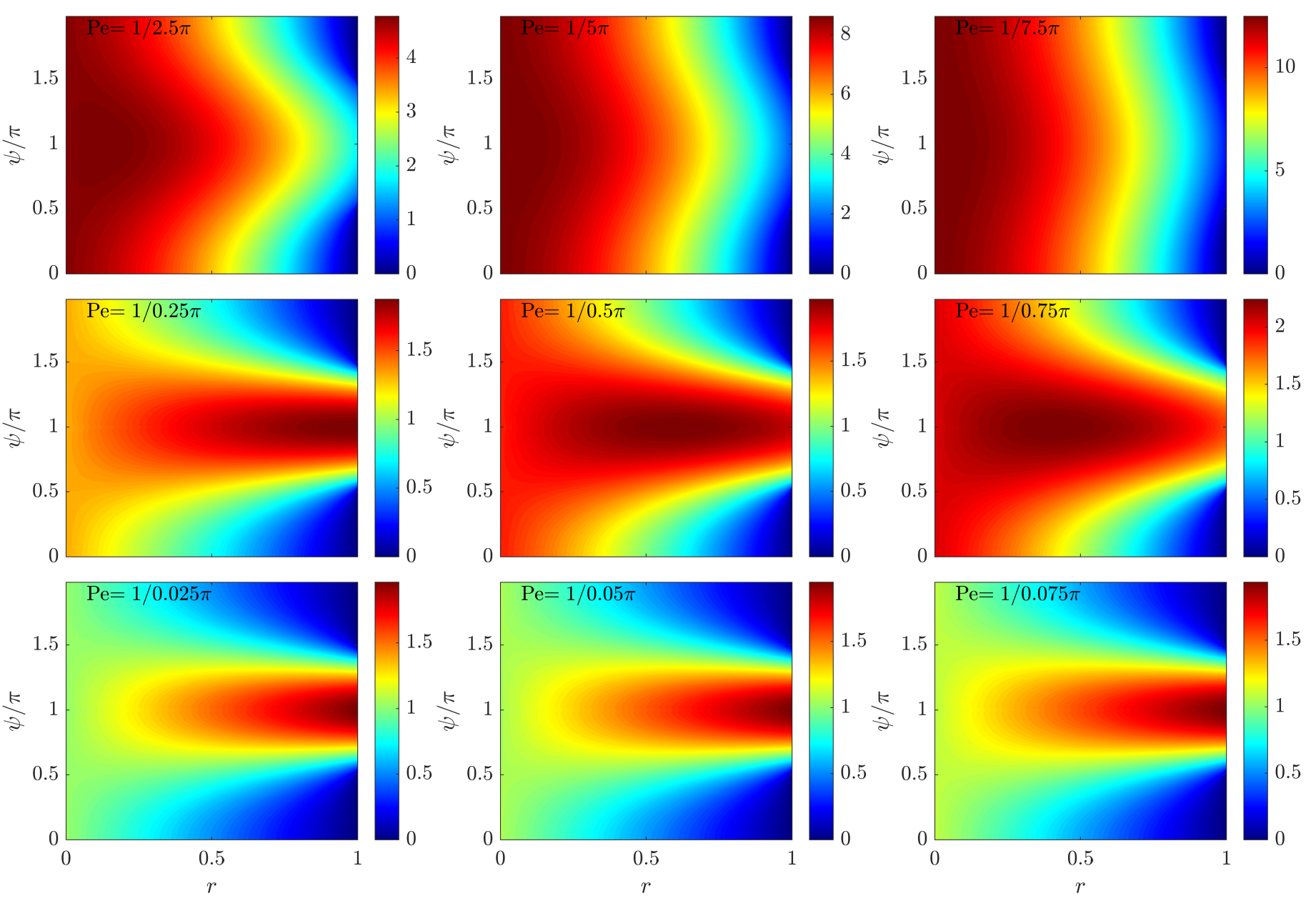}
    \caption{\label{fig:RTP disk MET contour} Contour plots of mean exit time of RTPs with different $\mathrm{Pe}$ in a flat disk. See the geometry in Fig. 3(a) of the main text.}  
\end{figure}

\begin{figure}[htbp]
    \centering
    \includegraphics[width=0.8\textwidth]{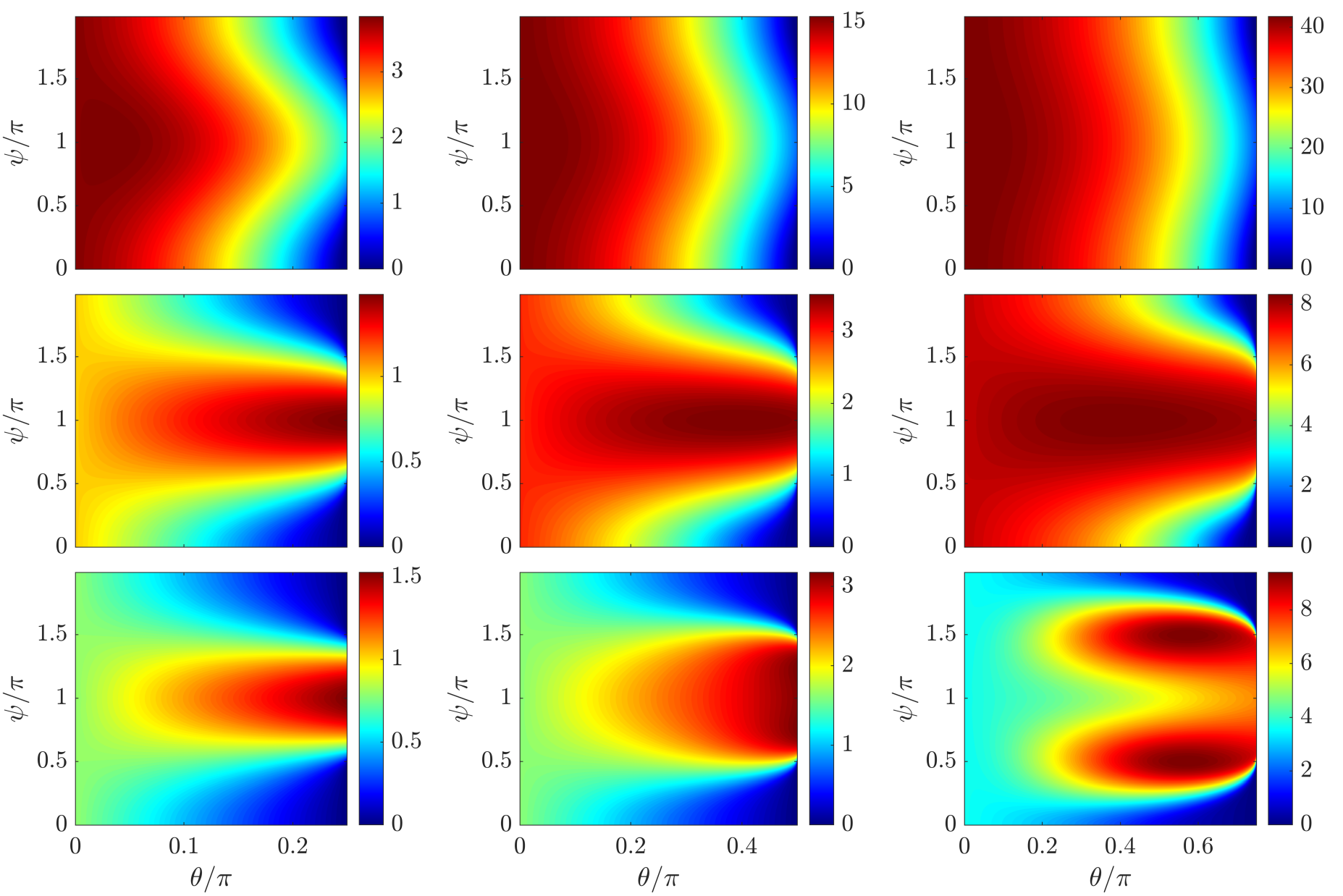}
    \caption{\label{fig:ABP MET sphere} Contour plots of mean exit time of ABPs with different $\mathrm{Pe}$and $\theta_0$ on a spherical cap. From top to bottom: $\mathrm{Pe} = 0.1,\;1,\;10$. From left to right: $\theta_0 = 0.25\pi,\; 0.5\pi,\; 0.75\pi$. See the geometry in Fig. 2 of the main text.}  
\end{figure}

\begin{figure}[htbp]
    \centering
    \includegraphics[width=0.8\textwidth]{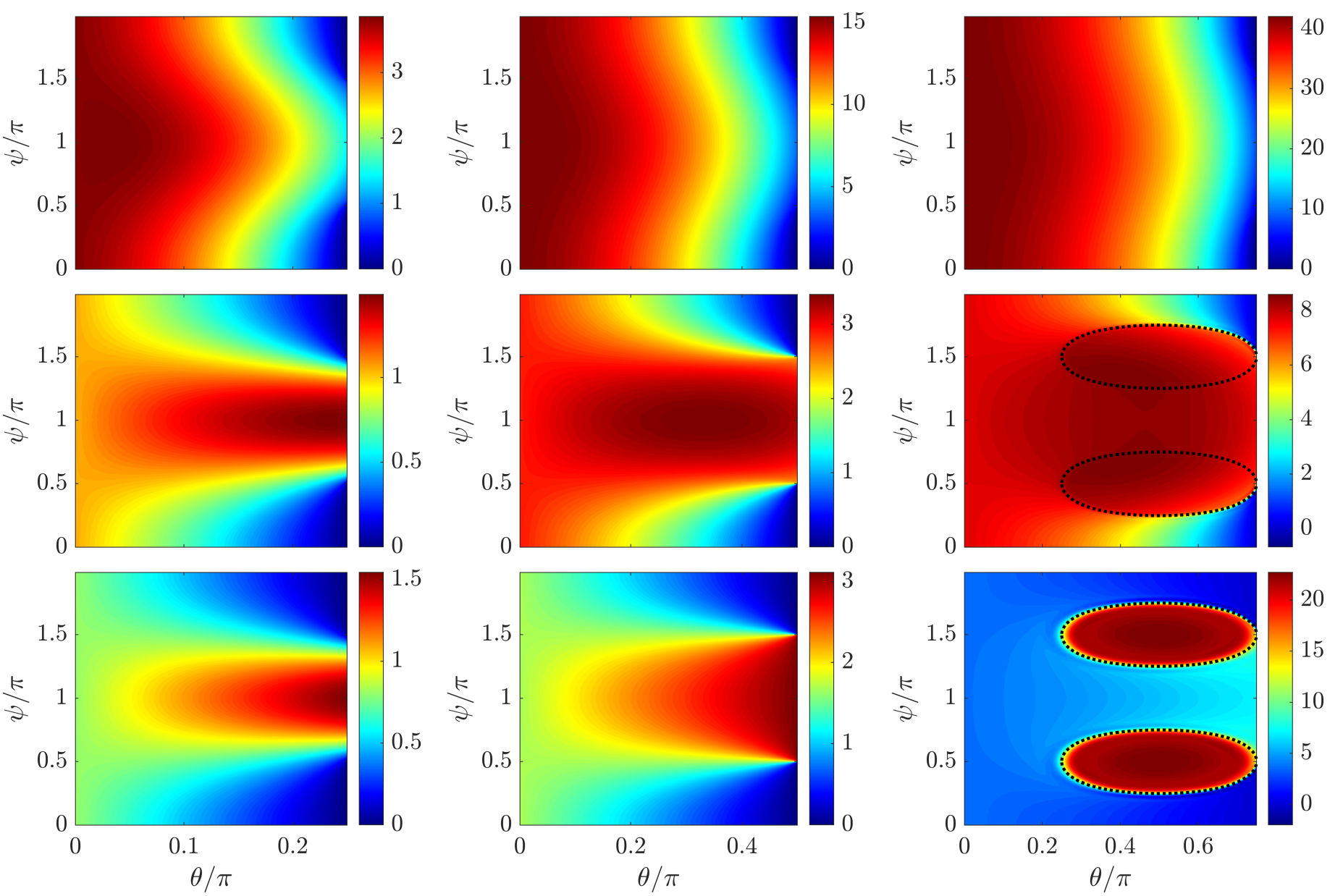}
    \caption{\label{fig:RTP MET sphere} Contour plots of mean exit time of RTPs with different $\mathrm{Pe}$ and $\theta_0$ on a spherical cap. From top to bottom: $\mathrm{Pe} = 0.1,\;1,\;10$. From left to right: $\theta_0 = 0.25\pi,\; 0.5\pi,\; 0.75\pi$. Dashed lines are the separatrix corresponding to $|\sin\psi \sin\theta| = \sin\theta_0$. See the geometry in Fig. 2 of the main text.}  
\end{figure}

\begin{figure}[htbp]
    \centering
    \includegraphics[width=\textwidth]{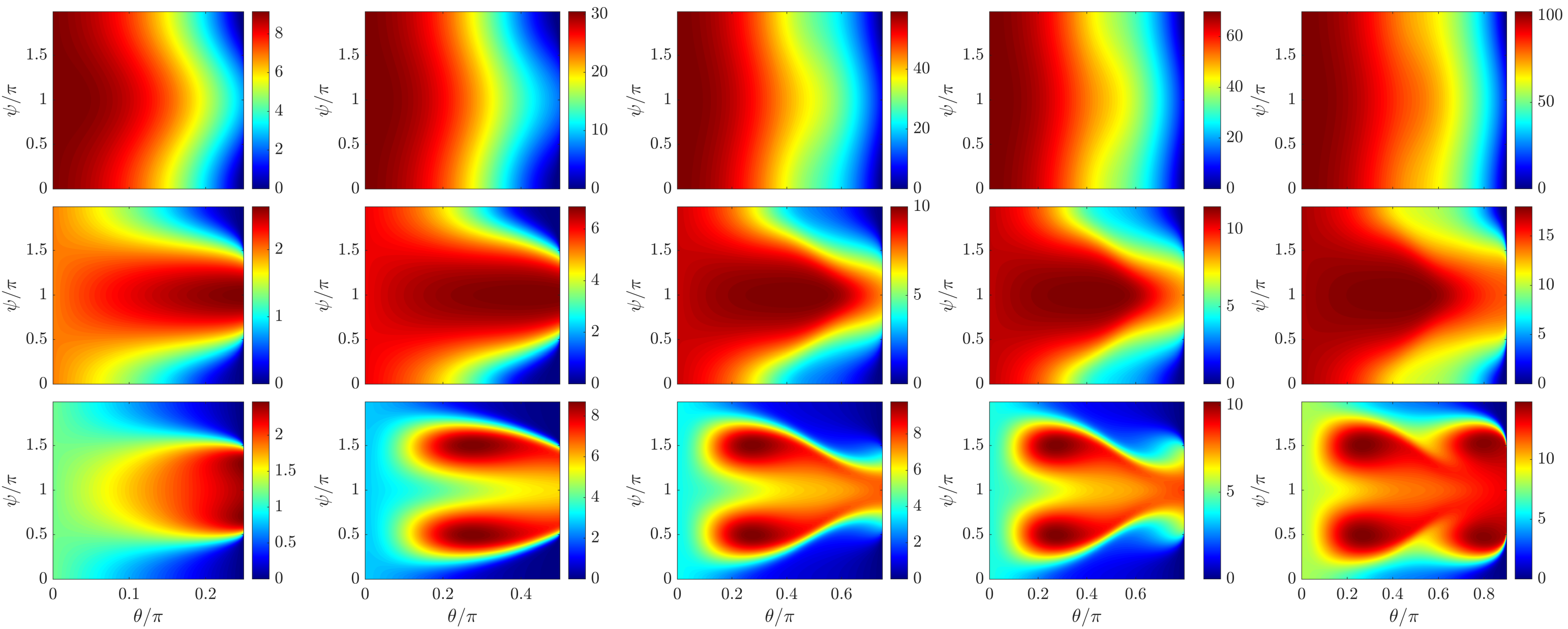}
    \caption{\label{fig:ABP MET deformed sphere} Contour plots of mean exit time of ABPs with different $\mathrm{Pe}$ and $\theta_0$ on an hourglass cap. From top to bottom: $\mathrm{Pe} = 0.1,\;1,\;10$. From left to right: $\theta_0 = 0.25\pi,\; 0.5\pi,\; 0.75\pi$. See the geometry in Fig. 2 of the main text.}  
\end{figure}

\begin{figure}[htbp]
    \centering
    \includegraphics[width=\textwidth]{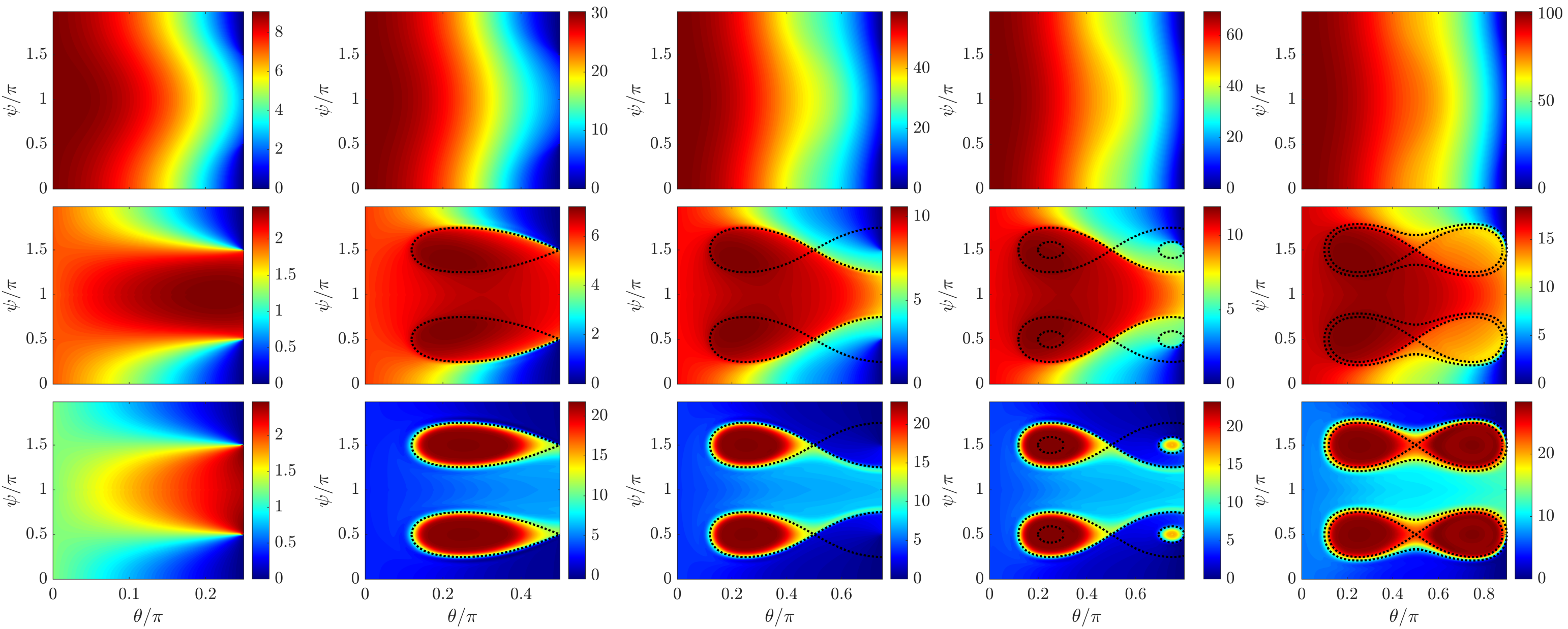}
    \caption[Contour plots of mean exit time of RTPs with different $\mathrm{Pe}$ and $\theta_0$ on an hourglass cap. From top to bottom: $\mathrm{Pe} = 0.1,\;1,\;10$. From left to right: $\theta_0 = 0.25\pi,\; 0.5\pi,\; 0.75\pi$. Dashed lines are the separatrix corresponding to $|R(\theta) \sin\psi \sin\theta| = R(\theta_0)\sin\theta_0$ and ]{\label{fig:RTP MET deformed sphere} Contour plots of mean exit time of RTPs with different $\mathrm{Pe}$ and $\theta_0$ on a hourglass cap. From top to bottom: $\mathrm{Pe} = 0.1,\;1,\;10$. From left to right: $\theta_0 = 0.25\pi,\; 0.5\pi,\; 0.75\pi$. Dashed lines are the separatrix corresponding to $|R(\theta) \sin\psi \sin\theta| = R(\theta_0)\sin\theta_0$ and $|R(\theta) \sin\psi \sin\theta| = R(\pi/2)$. See the geometry in Fig. 2 of the main text.}  
\end{figure}

Figures \ref{fig:ABP disk MET contour} to \ref{fig:RTP MET deformed sphere} show mean exit time fields in $(\theta,\psi)$ phase space for different choices of $\mathrm{Pe}$ and $\theta_0$ for ABPs and RTPs on: a flat disk (\ref{fig:ABP disk MET contour} and \ref{fig:RTP disk MET contour}), a spherical cap (\ref{fig:ABP MET sphere} and \ref{fig:RTP MET sphere}), and an hourglass shape (\ref{fig:ABP MET deformed sphere} and \ref{fig:RTP MET deformed sphere}).  

\section{Supplementary video captions}

\paragraph{Movie S1}  Monte Carlo particle simulations showing both ATPs and RTPs on a spherical cap with $\mathrm{Pe}=10$ and $\theta_0=0.75\pi$.  

\paragraph{Movie S2}  Monte Carlo particle simulations showing both ATPs and RTPs on an hourglass-shaped surface with $\mathrm{Pe}=10$ and $\theta_0=0.8\pi$.

\newpage

\normalem
\bibliographystyle{unsrt}
\bibliography{apssamp.bib}